\documentclass[10pt,aps,prd,showpacs,nofootinbib,twocolumn]{revtex4-2}

\usepackage{amssymb,amsmath,amsfonts}
\usepackage[usenames]{color}
\usepackage[pdftex]{graphicx}
\usepackage[english]{babel}
\usepackage{amsmath,amssymb}
\usepackage{multirow}
\usepackage{txfonts}
\usepackage[percent]{overpic}
\usepackage{overpic}
\usepackage{subfig}

\begin{document}

\title{Equation-of-state-insensitive measure of neutron star stiffness}

\author{Jayana A. Saes}
\email{jayanasaes@id.uff.br}
\affiliation{Instituto de F\'isica, Universidade Federal Fluminense, Niter\'oi, Rio de Janeiro, 24210-346, Brazil.}
\author{Raissa F.\ P.\ Mendes}
\email{rfpmendes@id.uff.br}
\affiliation{Instituto de F\'isica, Universidade Federal Fluminense, Niter\'oi, Rio de Janeiro, 24210-346, Brazil.}

\date{\today}

\begin{abstract}
Universal relations (i.e., insensitive to the equation of state) between macroscopic properties of neutron stars have proven useful for a variety of applications---from providing a direct means to extract observables from data to breaking degeneracies that hinder tests of general relativity. Similarly, equation-of-state-insensitive relations directly connecting macroscopic and microscopic properties of neutron stars can potentially provide a clean window into the behavior of nuclear matter. In this work, we uncover a tight correlation between certain macroscopic properties of a neutron star---its compactness $C$, moment of inertia $\bar{I}$ and tidal deformability $\bar{\Lambda}$---and the ratio $\alpha_c \equiv p_c/\epsilon_c$ of central pressure to central energy density, which can be interpreted as a mean notion of the stiffness of nuclear matter inside that object. We describe interesting properties of this stiffness measure, quantify the (approximate) universality of the $\alpha_c - C/\bar{I}/\bar{\Lambda}$ relations, and explore its consequences in the face of recent and future neutron star observations.
\end{abstract}


\maketitle

\section{Introduction}

Different observational channels are nowadays converging to compose an increasingly clear picture of neutron stars (NSs). Radio observations of rotation-powered pulsars have enabled accurate mass measurements for the components of a number of binary systems (see, e.g.~\cite{Kramer2006,Weisberg2010,Fonseca2014}), while pulse profile modeling of the x-ray data from the Neutron-Star-Interior-Composition Explorer (NICER) already provided measurements of the radius of two NSs \cite{Miller2019,Riley2019,Miller2021,Riley2021}, improving the accuracy of previous radii estimates. Additionally, observation of gravitational waves (GWs) from the binary NS merger GW170817 allowed a first measurement of the tidal deformability \cite{GW170817,GW170817eos}, with its electromagnetic counterpart providing valuable information about the remnant \cite{multimessenger}. 
Future years should witness the measurement of additional NS properties, such as their moment of inertia \cite{Kramer2009} or characteristic pulsation modes \cite{Clark2014,Yang2018,Torres-Rivas2019}.

All of these observables depend on the nuclear equation of state (EOS), which, for a cold NS, consists of a one-parameter relation between pressure and rest-mass density, $p = p(\rho)$. At the supranuclear densities thought to exist in NS cores, the EOS is mostly unconstrained by nuclear physics experiments. As a result, different theoretical strategies to extrapolate nuclear physics models to this high density regime give rise to disparate predictions for NS properties \cite{Lattimer2016,Ozel2016}. 

Interestingly, EOS-insensitive relations have been found to exist between several macroscopic NS observables (see e.g.~\cite{Yagi2017} and references therein). These relations are particularly useful to break degeneracies that would otherwise hinder an accurate estimation of NS properties. For example, the ``binary Love relation'' \cite{Yagi2016,Yagi2016a} between certain combinations of the tidal deformabilities of the components of a binary system can be used to infer the individual deformabilities, which are not currently measurable in GW data. Moreover, EOS-insensitive relations make comparison between various measurements of NS properties more straightforward. For instance, one can use the Love-$C$ \cite{Yagi2013a,Yagi2013} relation, between the tidal deformability and the NS compactness, to translate a measurement of mass and tidal deformability---inferred from GW data---into an estimate of the NS radius, making comparison with other radius measurements (say, from NICER data) more direct. Some EOS-insensitive relations, such as those connecting moment of inertia, tidal deformability and quadrupole moment (``I-Love-Q'') are incredibly tight, holding to roughly percent accuracy \cite{Yagi2013a}. 

EOS-insensitive relations have also been found, relating directly macroscopic and microscopic observables. For instance, pressure at $\sim$1-2 times the nuclear saturation density ($\rho_\text{sat} = 2.8 \times 10^{14}$ g cm${}^{-3}$) was found to correlate with the radius and tidal deformability of a typical NS \cite{Lattimer2001,Lim2018}, while pressure at higher densities ($\sim$7-8 $\rho_\text{sat}$) is thought to determine their maximum mass \cite{Ozel2009}. Estimates for the pressure at fiducial densities, coming from NS observations, can then be fed into nuclear physics models, constraining properties of the nuclear interaction such as the nuclear symmetry energy \cite{Lattimer2014,Drischler2021}.

The aim of this work is to introduce and explore a new approximately universal relation between certain macroscopic NS properties and a microscopic measure of the mean stiffness of nuclear matter inside a NS, namely, the ratio $\alpha_c \equiv p_c/\epsilon_c$ of central pressure to central energy density. This stiffness measure harbors similarities with more commonly used notions, such as the maximum speed of sound inside a NS. However, it correlates much more strongly with macroscopic observables such as the NS compactness ($C$), dimensionless moment of inertia ($\bar{I}$) or tidal deformability ($\bar{\Lambda}$). The approximately universal $\alpha_c-C/\bar{I}/\bar{\Lambda}$ relations are analyzed in this work both for a restricted set of realistic EOS and for a larger set of $\sim 60,000$ phenomenological EOS, in two different EOS parametrizations \cite{OBoyle2020,spectrallindblom}. Our main results are condensed in Figs.~\ref{fig:universal_realistic} and \ref{fig:universal_gpp}. As an example, the $\alpha_c-\bar{\Lambda}$ relation is found to hold to a maximum error of $\sim 8\%$ for the set of realistic EOS, and to a maximum error of $\sim 38\%$ for the set of (generalized) piecewise-polytropic EOS, with 90\% of the error below $\sim 7\%$ in this case. Measurements of tidal deformability, moment of inertia or compactness can thus be translated into estimates for $\alpha_c$, providing direct information about the behavior---and extremeness---of nuclear matter inside NSs.

This work is organized as follows. We begin in Sec.~\ref{sec:properties} by describing the properties and physical interpretation of the stiffness measure $\alpha_c$. Section \ref{sec:EOSspace} contains the description of the set of realistic and phenomenological EOS used subsequently. Our main results for the approximate universality of the $\alpha_c-C/\bar{I}/\bar{\Lambda}$ relations are presented in Sec.~\ref{sec:universality}, while Sec.~\ref{sec:observations} explores how present and future observations may constrain this microscopic quantity. Section \ref{sec:conclusions} gathers our main conclusions. We adopt natural units, $c = G = 1$, unless stated otherwise.

\section{Properties of the stiffness measure} \label{sec:properties}

\begin{figure}[t]
    \includegraphics[width=0.4 \textwidth]{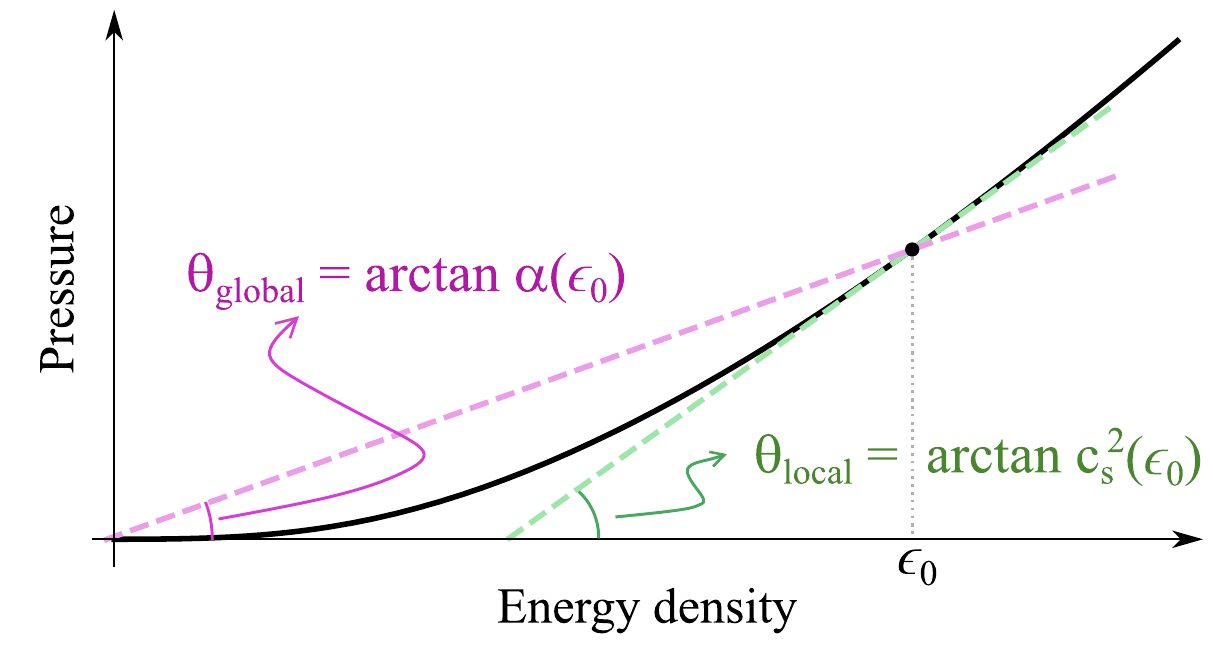}
   \caption{Geometrical interpretation of $\alpha$ (resp., $c_s^2$) as a global (resp., local) notion of the stiffness of nuclear matter. While $c_s^2(\epsilon_0)$ governs the change in pressure caused by an infinitesimal increment in energy density around $\epsilon = \epsilon_0$, $\alpha(\epsilon_0)$ determines the total increase in pressure from $\epsilon \approx 0$ (at the stellar surface) up to $\epsilon = \epsilon_0$ (say, at the stellar center).}
   \label{fig:illustration}
\end{figure}

The stiffness of nuclear matter encodes how pressure increases as density increases. Commonly used measures include the adiabatic sound speed, 
\begin{equation}\label{eq:soundspeed}
    c_s = \sqrt{\left. \frac{\partial p}{\partial \epsilon} \right|_{s}}
\end{equation}
and the adiabatic index, 
\begin{equation} \label{eq:adindex}
    \Gamma_1 = \left( \frac{\partial \ln p}{\partial \ln \rho} \right)_s,
\end{equation}
where $p$, $\epsilon$, $\rho$, and $s$ are the fluid pressure, energy density, rest-mass density, and entropy as measured by comoving observers. Both $c_s$ and $\Gamma_1$ are local stiffness measures: For each value of the energy density, $c_s^2(\epsilon)$ measures the slope of the $p(\epsilon)$ relation, and similarly for $\Gamma_1$. On the other hand, one can think of the ratio
\begin{equation}\label{eq:alpha}
    \alpha(\epsilon) \equiv \frac{p(\epsilon)}{\epsilon},
\end{equation}
when evaluated at the stellar center [$\alpha_c \equiv \alpha(\epsilon_c)$], as a global, or mean, notion of the stiffness of nuclear matter inside that NS. Indeed, since pressure vanishes at the stellar surface ($p_s=0$) and $\epsilon_s/\epsilon_c \approx 0$, $\alpha_c \approx (p_c - p_s) /(\epsilon_c - \epsilon_s)$, the ratio between the total increase in pressure and the total increase in energy density from the surface to the center. This is illustrated in Fig.~\ref{fig:illustration}.

Similarly to the sound speed, the ratio $p/\epsilon$ is deeply connected to the principle of causality, which requires that the velocity of causal influences should not exceed the speed of light. 
For a perfect fluid, with energy-momentum tensor
\begin{equation} \label{eq:perfectfluid}
    T^{\mu\nu} = (\epsilon + p) u^\mu u^\nu + p g^{\mu\nu},
\end{equation}
where $u^\mu$ is the four-velocity of fluid elements, the adiabatic sound speed determines the propagation speed of sound waves, and causality then implies that \cite{Anile1990}
\begin{equation}\label{eq:clessthan1}
    c_s \leq 1.
\end{equation}

On the other hand, the dominant energy condition (DEC)---which is the statement that the current of energy-momentum density, $-T^\mu_{\,\,\nu} \xi^\nu$, as measured by observers with four-velocity $\xi^\mu$, should be future-directed or null---implies, for a perfect fluid \eqref{eq:perfectfluid}, that $\epsilon \geq |p|$, or
\begin{equation} \label{eq:alphalessthan1}
    \alpha \leq 1.
\end{equation}
For a fluid characterized by a conserved energy-momentum tensor, the DEC implies that matter cannot travel faster than light, as it can be shown that if $T^{\mu\nu}$ vanishes on a close, achronal set $S$, it also vanishes in the domain of dependence of $S$ (see lemma 4.3.1 in \cite{Hawking1973}).
Therefore, condition \eqref{eq:alphalessthan1} is also a statement about causal behavior of the fluid\footnote{Note that it is in principle possible to construct special Lorentz invariant (causal) theories that violate both Eqs.~\eqref{eq:clessthan1} and \eqref{eq:alphalessthan1} \cite{Bludman1968,Bludman1970}.}.

Of course, since in the nonrelativistic limit nuclear matter satisfies $p \ll \epsilon \approx \rho$, for the bound $\alpha = 1$ to be achieved, the fluid must first go superluminal; Eq.~\eqref{eq:clessthan1} is therefore more restrictive than \eqref{eq:alphalessthan1}. In particular, it is interesting to consider the case of an EOS that is ``maximally soft'', $p(\epsilon) = 0$, for $\epsilon \leq \epsilon_0$, and ``maximally stiff'', $p(\epsilon) = \epsilon - \epsilon_0$, for $\epsilon \geq \epsilon_0$, which yields the most compact NS models \cite{Koranda1997}. The maximally compact configuration, with $C \approx 0.35$, has $\epsilon_c = 3.024 \epsilon_0$ and $p_c = 2.034 \epsilon_c$ and therefore
\begin{equation} \label{eq:alphamax}
    \alpha_c = 0.670.
\end{equation}
As we will see below, by requiring that the EOS satisfies Eq.~\eqref{eq:clessthan1}, the bound \eqref{eq:alphamax} is approached, rather than \eqref{eq:alphalessthan1}.

Another relevant value for the stiffness measure \eqref{eq:alpha} is $\alpha = 1/3$. Since $T \equiv g_{\mu\nu} T^{\mu \nu} = (3 \alpha - 1) \epsilon$ for a perfect fluid, $\alpha = 1/3$ corresponds to $T = 0$. Again, for the value $\alpha = 1/3$ to be reached inside a NS, the conformal bound $c_s^2 < 1/3$ \cite{Bedaque2015} must necessarily be violated in its interior. Notably, it has been pointed out before that the stellar compactness at which $T = 0$ at the stellar center is roughly EOS-independent \cite{Podkowka2018}. This can be seen as a particular case of the more general, EOS-insensitive relation between the NS compactness and any specific value of $\alpha_c$, which we now discuss.

\section{Space of equations of state} \label{sec:EOSspace}

In order to explore the relation between $\alpha_c$ and macroscopic properties of NSs, we consider both a set of 25 realistic EOS and a set of 60,000 phenomenological EOS following two different parametrization schemes. The realistic EOS are derived from nuclear physics models under a variety of approximation schemes (see Appendix A of Ref.~\cite{Abbott_2020} for references and details). They are all hadronic except for QHC19 \cite{Baym:2019iky}, which allows for a quark core. Here, we will use the parametrized form of these realistic EOS as described in Ref.~\cite{OBoyle2020} (cf.~Table III of that reference). The set of phenomenological EOS comprises $\sim 30,000$ (generalized) piecewise polytropic EOS with continuous speed of sound \cite{OBoyle2020} and $\sim 30,000$ spectral EOS \cite{spectrallindblom}. In this section we briefly describe these parametrizations, as well as the properties of the final set of EOS.

In the generalized piecewise polytropic (GPP) parametrization developed in Ref.~\cite{OBoyle2020}, one defines density intervals  $\rho_0<\rho_1<\rho_2<...$ above a certain value $\rho_0$, in such a way that
\begin{equation} \label{eq:pGPP}
p(\rho)=K_i\rho^{\Gamma_i} + \Lambda_i,\qquad \rho_{i-1}\leq\rho\leq\rho_i.
\end{equation}
The energy density $\epsilon(\rho)$ can be obtained from (\ref{eq:pGPP}) and the first law of thermodynamics. This generalizes the usual piecewise polytropic parametrization (see, e.g.~Ref.~\cite{Read2009}) by introducing the parameters $\Lambda_i$, which can be adjusted in order to guarantee continuity of the speed of sound. One disadvantage of this construction is that $\Gamma_i$ does not coincide with the adiabatic index (\ref{eq:adindex}) in each polytropic phase, rendering its interpretation less natural. 
Here, following Ref.~\cite{OBoyle2020}, we fix the crust to the SLy(4) EOS \cite{SLY}, and divide the core into three density intervals, with dividing densities $\rho_1 = 10^{14.87}$g/cm${}^3$ and $\rho_2 = 10^{14.99}$ g/cm${}^3$. Four free parameters define the EOS, which can be taken as $K_1$ and $\Gamma_i$, $i \in \{1,2,3\}$, while the remaining ones are determined by continuity and diferentiability requirements. 

We generate $\sim 30,000$ EOS by randomly sampling GPP parameters in the ranges $\Gamma_i \in [-2.0,8.0]$. For each value of $\Gamma_1$, $\log K_1$ is randomly sampled in the interval $[\log K_\text{min} (\Gamma_1), \log K_\text{max} (\Gamma_1)]$, where $K_\text{min/max}$ are determined by the requirement that the density $\rho_0$ that divides crust and core satisfies $ \rho_\text{SLy} < \rho_0 < \rho_1$, where $\rho_\text{SLy} = 5.317\times 10^{11}$ g/cm$^3$ is the last dividing density for a GPP parametrization of the crust (see Table II of Ref.~\cite{OBoyle2020}). We further require that all accepted EOS (i) are causal, in the sense that Eq.~(\ref{eq:clessthan1}) holds for all hydrodynamically stable configurations, and (ii) predict a maximum NS mass of at least $2.0 M_\odot$, allowing for systems such as J0740+6620 \cite{Fonseca2021}. We further assess the impact of setting an upper limit on the tidal deformability of a $1.4 M_\odot$ NS, adopting the GW170817 constraint (iii) $\bar{\Lambda}_{1.4 M_\odot}< 800$ \cite{GW170817}.
For each accepted EOS, we generate 50 equilibrium configurations with central densities between $\rho_\text{SLy}$ and that corresponding to the solution with maximum mass ($\rho_\text{max}$). For the analyses in the following section, we restricted attention to configurations with $C>0.05$.

In order to analyze the effect of a change to the EOS parametrization, we also consider the spectral parametrization presented in Ref.~\cite{spectrallindblom}. In this set-up the EOS is obtained as a solution to the differential equation
\begin{equation}
\label{eq:eqdifspectral}
    \frac{d\epsilon(p)}{dp} = \frac{\epsilon+p}{p\,\Gamma_1(p)},
\end{equation}
where the adiabatic index  $\Gamma_1(p)$ is expanded as 
\begin{equation} \label{eq:adiabatic_spectral}
    \Gamma_1(p)=\exp{\left[\sum_{k=0}^3 \gamma_k\, \left[\log{\left(\frac{p}{p_0}\right)}\right]^k \right]}.
\end{equation}\\
Here $\gamma_k$ are free parameters specifying the EOS and $p_0$ is a reference pressure that determines the initial condition $\epsilon_0 \equiv \epsilon (p_0)$ for the integration of Eq.~(\ref{eq:eqdifspectral}). It is defined as the pressure at the dividing density, chosen as  $\rho_0 = 2\times 10 ^{14}$ g/cm$^3$, between the low density EOS, which we take to be SLy \cite{SLY}, and the high density (core) EOS. The pressure at $\rho_0$ can be found from the definition (\ref{eq:adindex}) of the adiabatic index.

\begin{figure}[t]
    \centering
    \includegraphics[width=0.48 \textwidth]{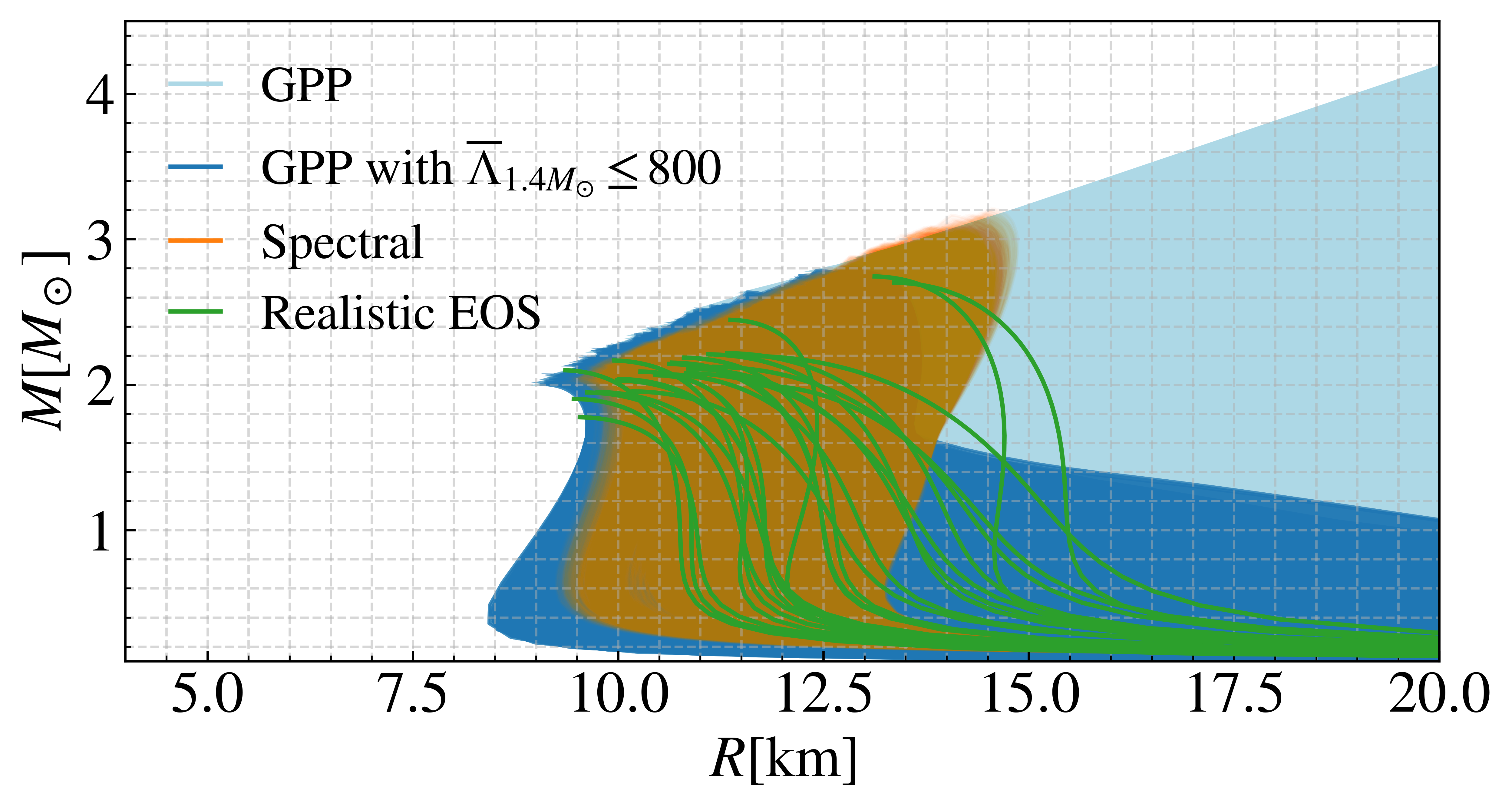}
    \caption{Mass-radius diagram for our set of 25 realistic EOS and 60,000 phenomenological EOS, of which half follow the GPP parametrization and half follow the spectral representation. The subset of GPP EOS obeying $\bar{\Lambda}_{1.4 M_\odot} < 800$ is highlighted to allow for a clearer comparison with the spectral set, for which this condition also holds.}
   \label{fig:massaraio}
\end{figure}

\begin{figure*}[t]
    \centering
    \includegraphics[width=\textwidth]{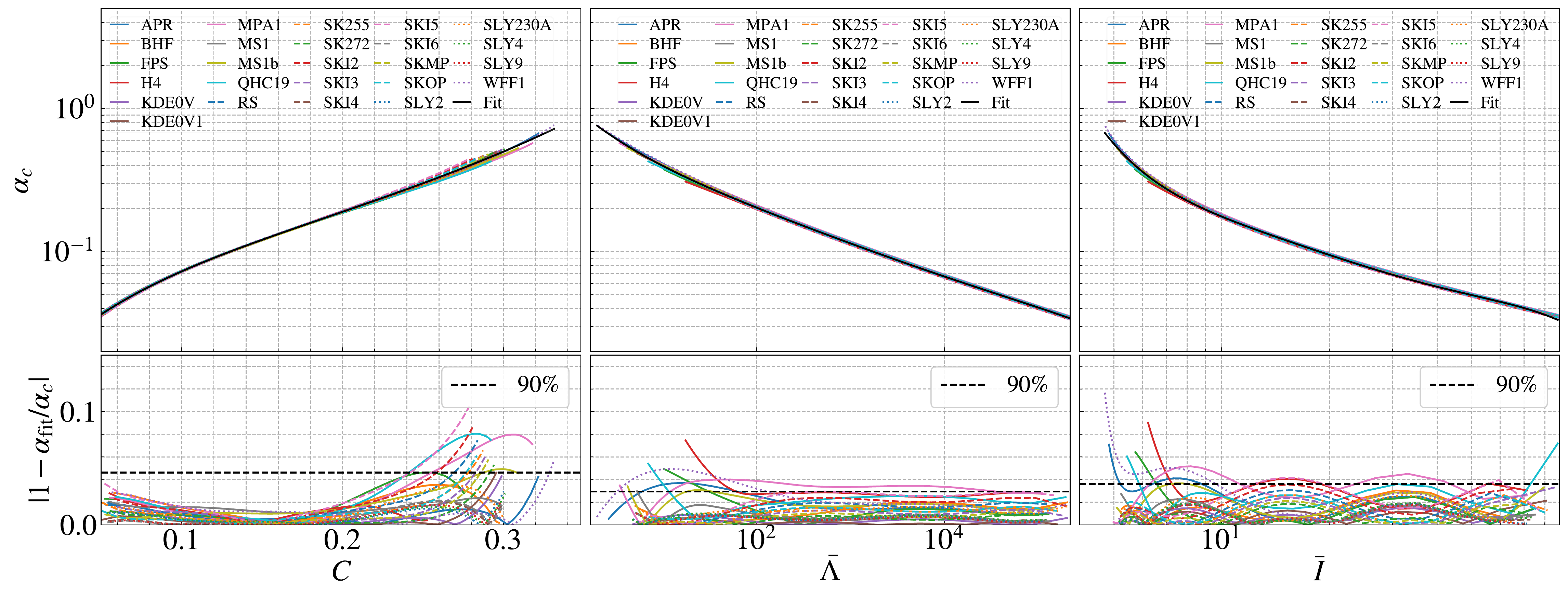}
   \caption{Approximately universal $\alpha_c - C/\bar{I}/\bar{\Lambda}$ relations for a set of 25 realistic EOS. A black line represents a fifth order polynomial fit [as in Eq.~(\ref{eq:fits})], and the bottom panels display the corresponding fractional error. Only configurations with $C>0.05$ are considered. }
   \label{fig:universal_realistic}
\end{figure*}

\begin{table*}[htb]
\centering
\begin{tabular}{c|c|c|c|c|c|c|c}
\hline \hline
EOS & Relation & $a_0$ & $a_1$ & $a_2$ & $a_3$ & $a_4$ & $a_5$  \\ \hline
Realistic & $\alpha_c - C$ &$ -4.3997$& $27.754$& $-144.38$& $541.86$& $-10.295\times10^2$& $857.16$\\ 
{}        & $\alpha_c - {\bar I}$ & $23.768$& $-38.842$& $24.44$& $-7.8292$& $1.245$& $-78.265\times10^{-3}$\\ 
{}        & $\alpha_c - \bar{\Lambda}$& $85.901\times10^{-3}$& $-58.019\times10^{-2}$ & $82.016\times10^{-3}$ & $-10.5109\times10^{-3}$& $6.967\times 10^{-4}$&$-1.8214\times 10^{-5}$\\ 
Phenomenological & $\alpha_c - C$  & $-4.619$ & $33.734$ & $-221.79 1$& $10.389\times10^{2}$ &$-24.278\times10^{-3}$ &2173.6 \\ 
(GPP)   & $\alpha_c - {\bar I}$ & $18.705$& $-29.445$& $17.655$& $-5.4557$& $84.167\times10^{-2}$& $-51.543\times10^{-3}$\\ 
{}        & $\alpha_c - \bar{\Lambda}$&$-77.943\times10^{-3}$& $-43.034\times10^{-2}$& $27.195\times10^{-3}$& $-16.601 \times 10^{-4}$& $4.2749 \times 10^{-5}$ & $-9.4932 \times 10^{-8}$ \\ \hline \hline
\end{tabular}
\caption{Fit coefficients to the $\alpha_c - C / \bar{I}/ \bar{\Lambda}$ relations, for a set of 25 realistic EOS (upper rows), and a set of 30,000 phenomenological (GPP) EOS (bottom rows). }
\label{tab:fit}
\end{table*}

In principle, one could generate spectral EOS by directly sampling the parameters $\gamma_k$. However, we have found this procedure to be inefficient, since the adiabatic index is very sensitive to small changes in $\gamma_k$ due to the exponential factor in Eq.~(\ref{eq:adiabatic_spectral}), easily giving rise to nonphysical EOS. Instead, we have performed a random-walk in the EOS space, starting from the SLy parameters $\gamma_k^{\mathrm{SLy}}$ that can be found in Ref.~\cite{spectrallindblom}. Parameters for EOS $i$ are such that $\gamma_k^{i}=\gamma_k^{i-1}+r^i_k\,\gamma_k^{\mathrm{SLy}}$, where $\gamma_k^{0}=\gamma_k^{\mathrm{SLy}}$ and $r^i_k$ are sampled uniformly in the interval $[-0.05,0.05]$. In sampling the space of spectral EOS, we enforced conditions (i)-(iii) above, requiring causality, a $2.0 M_\odot$ lower bound on the maximum mass, and an upper bound on the tidal deformability of a $1.4 M_\odot$ NS ($\bar{\Lambda}_{1.4 M_\odot}< 800$).
Following this procedure, we generated $\sim$ 30,000 EOS; for each one, 50 equilibrium configurations were computed from a minimum compactness of $\sim 0.05$ and up to the central density $\rho_\text{max}$ of the maximum-mass configuration.

Figure \ref{fig:massaraio} shows the final set of EOS. We see that, although the spectral set contains EOS with large maximum masses ($> 3.0 M_\odot$) and largely accommodates the behavior of realistic EOS, it is still centered around the SLy EOS. On the other hand, the random sampling of GPP parameters allows for a much larger region of the EOS space to be covered with less samples. In what follows, we will use the set of 30,000 GPP EOS as our standard set.

\begin{figure*}[t]
    \centering
    \includegraphics[width=\textwidth]{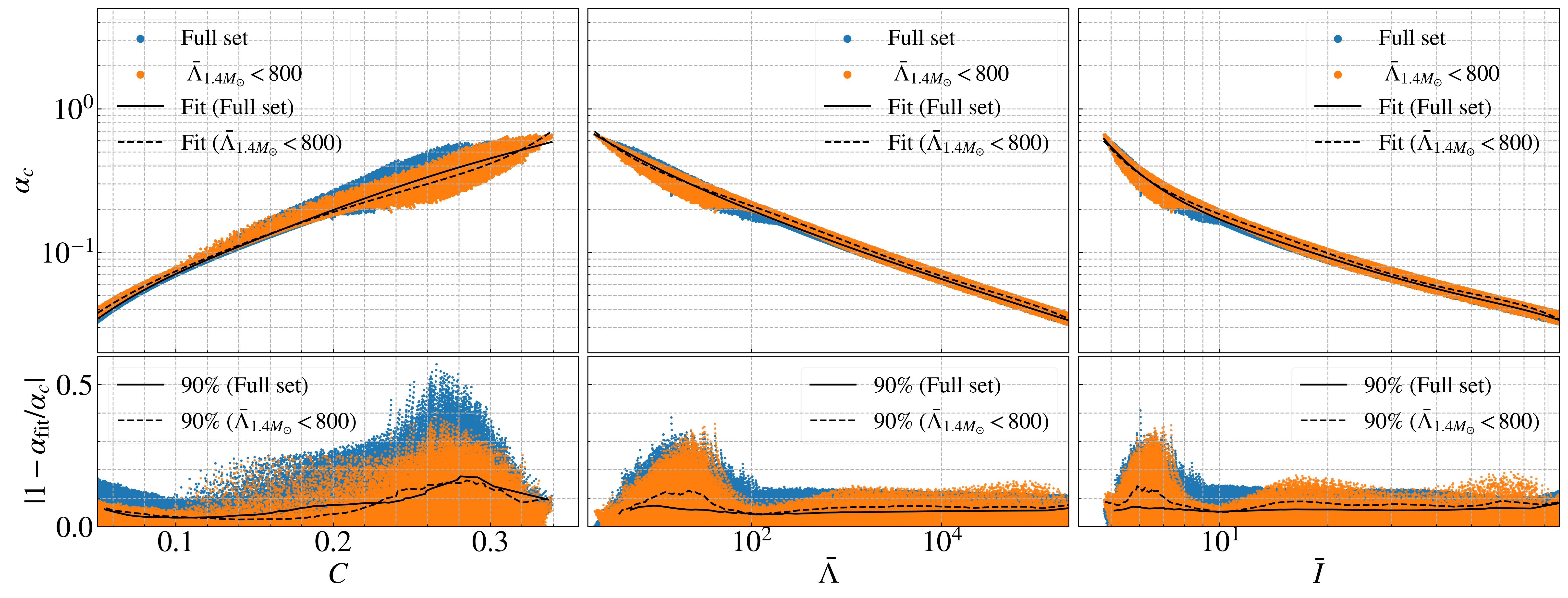}
   \caption{Approximately universal $\alpha_c - C/\bar{I}/\bar{\Lambda}$ relations for a set of $\sim$30,000 phenomenological EOS in the GPP parametrization (blue), and for the subset obeying the additional requirement that $\bar{\Lambda}_{1.4 M_\odot} < 800$ (orange). A solid (resp., dashed) black curve represents a fifth order polynomial fit [cf.~Eq.~(\ref{eq:fits})] to the full (resp., restricted) set of EOS. Bottom panels display the corresponding fractional error, with lines enclosing 90\% of the errors. Only configurations with $C>0.05$ are displayed.}
   \label{fig:universal_gpp}
\end{figure*}

\begin{figure*}[th]
    \centering
\includegraphics[width=\textwidth]{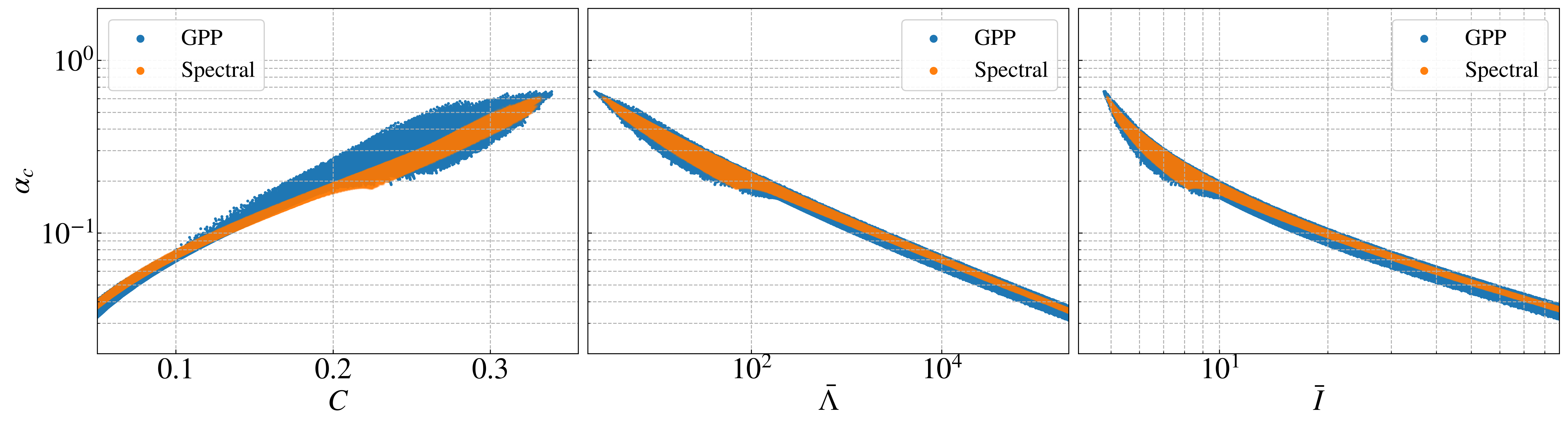}
   \caption{Universal $\alpha_c - C/\bar{I}/\bar{\Lambda}$ relations for a set of $\sim$30,000 phenomenological EOS with the GPP parametrization (blue), and for $\sim$30,000 EOS with the spectral parametrization (orange). Only configurations with $C>0.05$ are displayed.}
   \label{fig:universal_spectral}
\end{figure*}

\section{Approximate universality} \label{sec:universality}

Here we explore the relation between the stiffness measure $\alpha_c = p_c/\epsilon_c$ and the NS compactness, moment of inertia and tidal deformability, for different choices of realistic or phenomenological EOS. The NS compactness, $C = M/R$, is computed from its mass $M$ and radius $R$, obtained by numerically solving the TOV equations for spherically symmetric hydrostatic equilibrium. The NS moment of inertia, $I = J/\Omega$, is computed from its angular momentum $J$ and angular velocity $\Omega$ in the approximation of slow rotation \cite{Hartle1967}. From it we construct the dimensionless quantity ${\bar I} = I/M^3$, which is known to obey an EOS-insensitive relation with $C$ \cite{Breu2016}. Finally, the tidal deformability is computed by considering quadrupolar perturbations to an isolated NS; it measures the strength of the quadrupole moment $Q_{ij}$ induced in a NS by an external tidal field $\mathcal{E}_{ij}$: $Q_{ij} = - \Lambda \mathcal{E}_{ij}$ \cite{Hinderer2008,Damour2009,Binnington2009}. Again, we consider its dimensionless version, $\bar{\Lambda} = \Lambda / M^5$, which is also known to obey an EOS-insensitive relation with $C$ and $\bar{I}$ \cite{Maseli2013}.
In order to quantify the approximate universality of the $\alpha_c - C/\bar{I}/\bar{\Lambda}$ relations, we consider fifth-order polynomial fits of the form
\begin{equation}\label{eq:fits}
     \ln\alpha_{c} = \sum_{j = 0}^5 a_j \mathcal{O}^j,
\end{equation}
where $\mathcal{O} \in \{C, \ln {\bar I},\ln \bar{\Lambda}\}$. The coefficients $a_j$ for the set of realistic and phenomenological (GPP) EOS are given in Table \ref{tab:fit}.
Note that these fits can be optimized when some other property of a NS (such as its mass) is known.

Results for the set of 25 realistic EOS are shown in Fig.~\ref{fig:universal_realistic}, together with fits of the form (\ref{eq:fits}) and the corresponding fractional errors.
For low compactness ($C \lesssim 0.2$), the $\alpha_c - C$ relation holds to a relative error of less than $4\%$, but the error increases for higher values of $C$, up to a maximum of $\sim 11\%$. Similarly, the universality of the $\alpha_c - {\bar I}/\bar{\Lambda}$ relations is stronger for larger values of ${\bar I}$ and $\bar{\Lambda}$, and slightly weaker for lower values: The $\alpha_c - {\bar I}$ relation holds to a maximum error of $\sim 11 \%$, while the $\alpha_c - \bar{\Lambda}$ relation holds to a maximum error of $\sim 8\%$ for this set of realistic EOS. 

To further probe the approximate universality of the $\alpha_c - C/\bar{I}/\bar{\Lambda}$ relations, we explore a set of phenomenological EOS (cf.~Sec.~\ref{sec:EOSspace}).
Figure \ref{fig:universal_gpp} shows the $\alpha_c - C/ \bar{I}/ \bar{\Lambda}$ relations for a set of $\sim$ 30,000 GPP EOS, together with fits of the form (\ref{eq:fits}) and the corresponding fractional errors. The maximum value of $\alpha_c$ generated in this set of equilibrium solutions was $\sim 0.662$, lower than the bound (\ref{eq:alphamax}). 
For this set of EOS, the $\alpha_c - C$ relation holds to a maximum error of $\sim 57\%$, with 90\% of errors below $\sim 18\%$; the $\alpha_c - \bar{I}$ relation holds to a maximum error of $\sim 41\%$, with 90\% of errors below $\sim 8\%$; and the $\alpha_c - \bar{\Lambda}$ relation holds to a maximum error of $\sim 38\%$, with 90\% of errors below $\sim 8\%$. These errors are much lower than those displayed by other relations linking macroscopic and microscopic properties (as we discuss in Sec.~\ref{sec:universality_comparison} below), and are comparable with those exhibited by some universal relations between macroscopic observables (such as between the quadrupole and higher multipoles of the tidal deformability \cite{Yagi2014,Godzieba2021}).
It is worth mentioning that roughly 30\% of the GPP EOS in our set display an important softening at high densities, in such a way the speed of sound is not a monotonically increasing function of $\rho$ (up to $\rho_\text{max}$). These EOS contribute significantly to the larger spread seen in Fig.~\ref{fig:universal_gpp} for high values of $C$ (or low $\bar{I}$, $\bar{\Lambda}$). 

Finally, Fig.~\ref{fig:universal_spectral} compares the $\alpha_c - C/\bar{I}/\bar{\Lambda}$ relations for the set of GPP and spectral EOS.
The trend is similar, but a smaller region is covered by the set of spectral EOS. This is expected from the discussion in Sec.~\ref{sec:EOSspace} since, for a fixed number of EOS, a random sampling of GPP parameters allows for a larger diversity of EOS behaviors than the random walk performed in the space of spectral EOS.

\subsection{Comparison to other relations between macroscopic and microscopic quantities} \label{sec:universality_comparison}

To conclude this section, it is worthwhile to discuss how the $\alpha_c - C/\bar{I}/\bar{\Lambda}$ relations compare with other relations between macroscopic and microscopic NS properties.

\begin{figure}[th]
    \centering
    \includegraphics[width=0.45\textwidth]{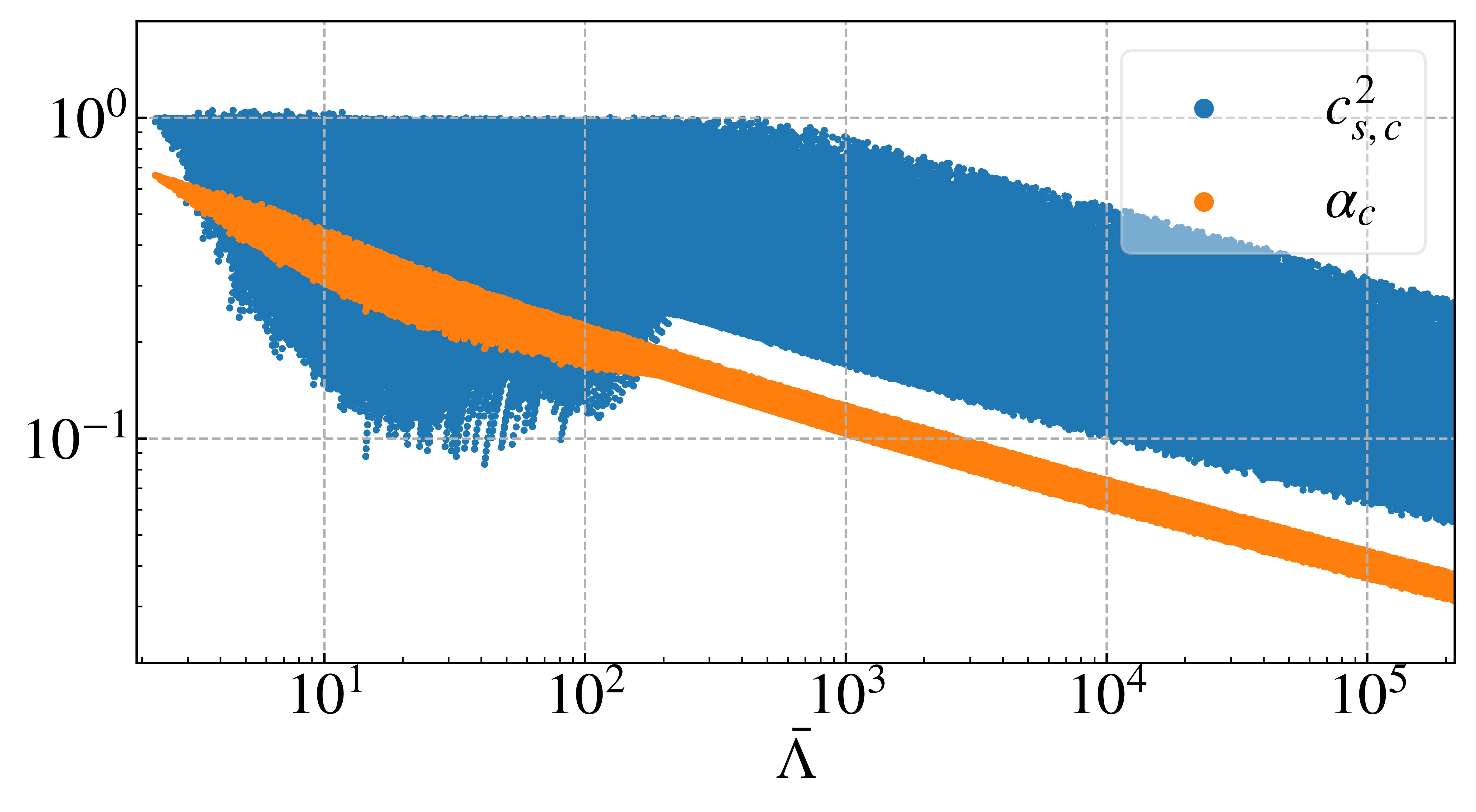}
   \caption{The ratio $\alpha_c = p_c/\epsilon_c$ and the square of the speed of sound at the stellar center, $c_{s,c}^2$, are shown as a function of $\bar{\Lambda}$ for a collection of equilibrium configurations obeying a set of $\sim 30,000$ phenomenological (GPP) EOS. }
   \label{fig:csL}
\end{figure}

First, it is interesting to stress that although $\alpha$ shares some properties with the speed of sound, as discussed in Sec.~\ref{sec:properties}, the correlation that is seen between $\alpha_c$ and $C/\bar{I}/\bar{\Lambda}$ is much weaker for, say, the sound speed at the stellar center ($c_{s,c}$). This is evident from Fig.~\ref{fig:csL}, where the $\alpha_c - \bar{\Lambda}$ and $c_{s,c}^2 - \bar{\Lambda}$ relations are shown. 

\begin{figure*}[htp]
    \centering
    \includegraphics[width=\textwidth]{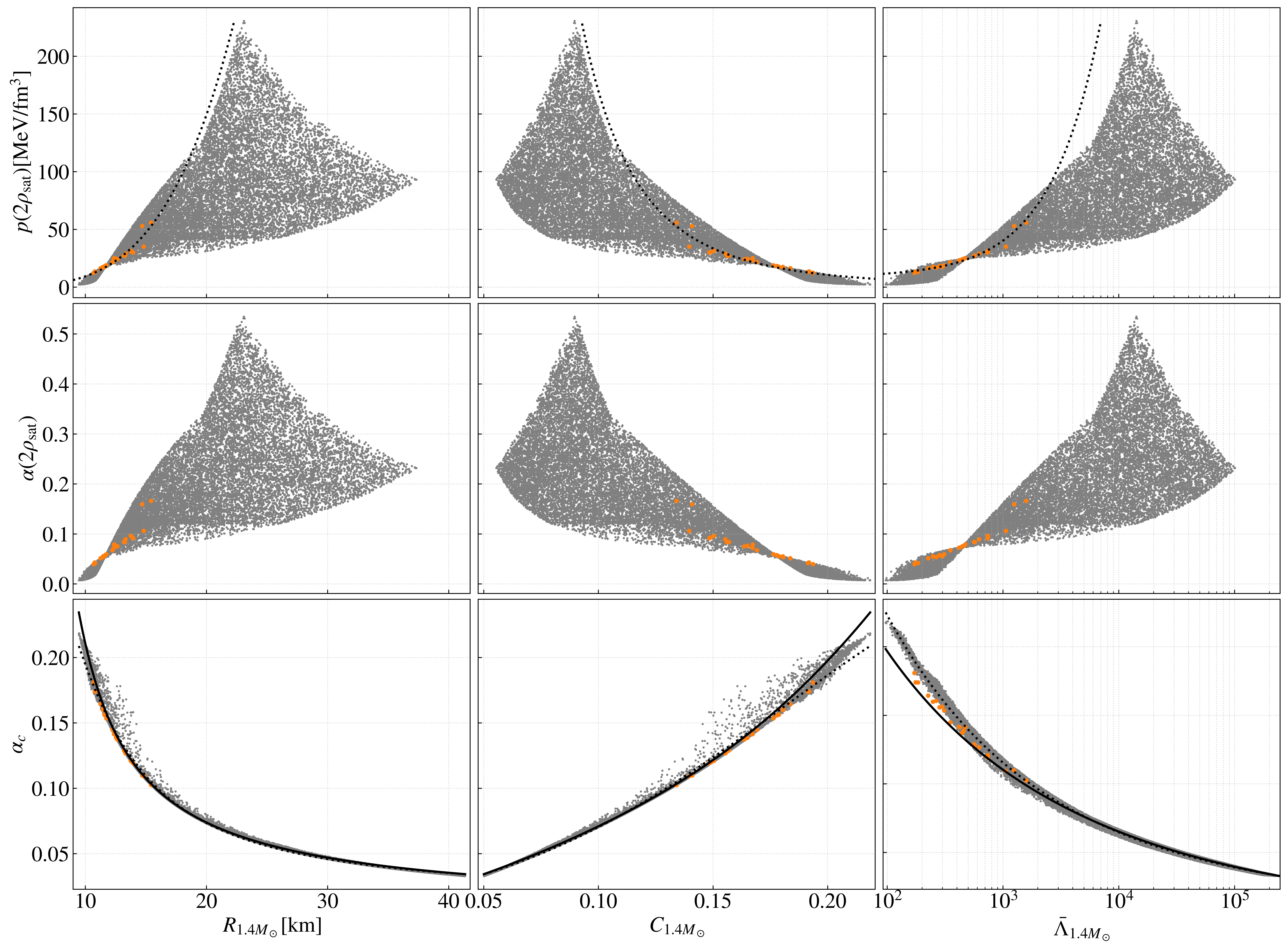}
   \caption{Top row: pressure at twice nuclear saturation density as a function of radius, compactness and dimensionless tidal deformability of a $1.4 M_\odot$ NS for a set of GPP (gray) and realistic (orange) EOS. A black dotted line represents the fitting formulas $p(2\rho_\text{sat})^{-1/4} R_{1.4 M_\odot} = 5.72$ km MeV$^{-1/4}$ fm$^{3/4}$ \cite{Lattimer2001} (first panel), 
   $p(2\rho_\text{sat})^{1/4} C_{1.4 M_\odot}= 0.361$ MeV$^{1/4}$ fm$^{-3/4}$, which is a rescaling of the first (second panel), and $\bar{\Lambda}_{1.4 M_\odot} = 31.59 \, p(2 \rho_\text{sat}) / (\text{MeV fm}^{-3}) - 272.36$ \cite{Lim2018} (third panel). Middle row: same as first row, but for $\alpha$ at twice nuclear saturation density. Bottom row: same as first row, but for $\alpha_c$. Solid black lines represent the fitting formulas from Table \ref{tab:fit}, derived for the full set of stellar configurations (in the GPP parametrization), while dotted lines represent second order polynomial fits for this restricted set (with $M = 1.4 M_\odot$); explicitly, they represent $\ln{\alpha_c}=-30.704\,C^2+18.959\,C -4.2432$ (from which one obtains the fitting formula for $R_{1.4 M_\odot}$) and $\ln{\alpha_c}=73.308\times10^{-4}\,\ln{\bar{\Lambda}^2} -0.37\,\ln{\bar{\Lambda}}+ 47.688\times10^{-3}$. Only configurations with $C > 0.05$ are shown.}
   \label{fig:comparison_lattimer}
\end{figure*}

Second, we compare the $\alpha_c - C/\bar{I}/\bar{\Lambda}$ relations with the EOS-independent relation uncovered by Lattimer and Prakash, who showed that the radius of a NS with fixed mass (or, equivalently, its compactness) correlates well with the pressure around nuclear saturation density \cite{Lattimer2001}. As a consequence of the Love-C (and I-C) relations \cite{Maseli2013,Breu2016}, a similar correlation is expected between the dimensionless tidal deformability (or moment of inertia) and pressure at 1-2 $\rho_\text{sat}$ (see, e.g.,~Ref.~\cite{Lim2018} for the case of the tidal deformability). This is illustrated for a $1.4 M_\odot$ NS in the first row of Fig.~\ref{fig:comparison_lattimer} for a set of GP (gray) and realistic (orange) EOS. The correlation is stronger for realistic EOS, but much weaker for the larger set of phenomenological EOS. 

Now, around nuclear saturation density, the ratio $\epsilon / \rho$ between energy density and rest mass density is not expected to differ appreciably from unity --- indeed, for the set of $\sim$30,000 GPP EOS, $1.03 \lesssim \epsilon(2 \rho_\text{sat}) / (2 \rho_\text{sat}) \lesssim 1.42$. As a consequence, a similar correlation is expected between $\alpha(1$-$2 \rho_\text{sat})$ and the radius (or compactness, tidal deformability and moment of inertia) of a NS with fixed mass. This is shown in the second row of Fig.~\ref{fig:comparison_lattimer}, which essentially mimics the first. 

One might thus wonder whether the $\alpha_c - C/\bar{I}/\bar{\Lambda}$ relations we report in our work might be an extrapolation of the well-known relation of Lattimer and Prakash for the higher densities present in the NS core. The plots on the third row of Fig.~\ref{fig:comparison_lattimer} show, on the other hand, that the correlation between $\alpha$ and $R_{1.4 M_\odot}$, $C_{1.4 M_\odot}$ and $\bar{\Lambda}_{1.4 M_\odot}$ is much stronger at the stellar center than around $\rho_\text{sat}$. Thus, it seems more robust to suggest that the  $\alpha_c - C/\bar{I}/\bar{\Lambda}$ relations are more fundamental --- in the sense of a weaker dependence on the EOS --- and universality deteriorates when extended to lower densities. Still, it is worth noticing that x-ray observations and GW170817 constraints favor values of $R_{1.4 M_\odot} < 15$ km, for which the scatter in the relations to $\alpha_c$ is somewhat larger.

\section{Connecting to observations} \label{sec:observations}

The nearly universal relations presented in this work enable a relatively precise determination of the pressure to energy-density ratio at the core of a NS given a precise measurement of either $C$, $\bar{\Lambda}$ or $\bar{I}$. 
To demonstrate this, in this section we determine the posterior distribution for $\alpha_c$ for three NSs with recently measured properties: The primary and secondary components of event GW170817, for which the tidal deformability has been measured \cite{GW170817eos}, and the massive pulsar J0740+6620, with mass $2.08 \pm 0.07 M_\odot$ (68.3\% credibility) \cite{Fonseca2021}, which recently had its radius measured \cite{Miller2021,Riley2021}.
Although NICER observations of pulsar J0030+0451 have also enabled estimates of its mass and radius \cite{Riley2019,Miller2019}, we do not consider this NS in our analysis, since its $\sim 1.4 M_\odot$ mass falls in the same range as the binary components that originated GW170817.
Next, we explore how these posteriors would change given more precise measurements of NS properties. 

Instead of basing our analysis on the fitting formulas presented in Table~\ref{tab:fit}, here we perform a full Bayesian analysis, as we describe below. The reason not to use the fitting formulas---especially in the case of broad experimental likelihoods---is that they do not carry information about the underlying EOS distribution, which may have little or no support in some regions of parameter space. 

In what follows, we compute $p(\alpha_c | \vec{D}, I)$, the posterior probability for $\alpha_c$ given some measurement of NS properties $\vec{D}$ and background information $I$, by marginalizing $p(\alpha_c , \vec{\theta}| \vec{D}, I)$  over EOS parameters (in the GPP parametrization) $\vec{\theta} = \{\log K_1, \Gamma_1, \Gamma_2, \Gamma_3\}$. In practice, the marginalization is performed via Monte Carlo integration,
$$
p(\alpha_c | \vec{D}, I) = \int d\vec{\theta} p(\alpha_c, \vec{\theta} | \vec{D},I) \approx \frac{V_\text{EOS}}{N} \sum_{i=1}^N p(\alpha_c,\vec{\theta}_i| \vec{D}, I),
$$
and we use as the set of $N$ randomly sampled EOS parameters $\vec{\theta}_i$ the same set generated for the previous analyses. The ranges for these parameters were described in Sec.~\ref{sec:EOSspace} and define the volume $V_\text{EOS}$.

The  probability $p(\alpha_c , \vec{\theta}| \vec{D}, I)$ that a star with properties $\vec{D}$ has a given value $\alpha_c$ of central pressure to central energy density and EOS parameters $\vec{\theta}$ is computed from Bayes theorem,
\begin{equation}
    p(\alpha_c,\vec{\theta} | \vec{D}, I) = \mathcal{N} p(\vec{D} | \alpha_c, \vec{\theta}, I) p(\alpha_c, \vec{ \theta} | I),
\end{equation}
where $\mathcal{N}$ is a normalization factor that can be determined \textit{a posteriori}. The prior $p(\alpha_c, \vec{\theta} | I)$ can be factored as 
$p(\alpha_c, \vec{\theta} | I) = p (\alpha_c | \vec{\theta}, I) p(\vec{\theta} | I)$.
We assume the prior $p(\vec{\theta} | I)$ on the GPP EOS coefficients to be uniform in the ranges discussed in Sec.~\ref{sec:EOSspace}, and $p (\alpha_c | \vec{\theta}, I)$ to be uniform in the range $\alpha(\rho_\text{SLy}) \leq \alpha_c \leq \alpha(\rho_\text{max})$, where $\rho_\text{max}$ is the central density for the most massive NS predicted by an EOS with parameters $\vec{\theta}$. The likelihood $p(\vec{D} | \alpha_c, \vec{\theta}, I)$ is either inferred from publicly available experimental distributions or assumed to be Gaussian. In computing this likelihood, one has to take into account the fact $\alpha(\rho)$ may not be a monotonically increasing function of $\rho$ in the range of interest, so that the inverse relation may not be single-valued.

\begin{figure}[th]
    \centering
    \includegraphics[width=0.45\textwidth]{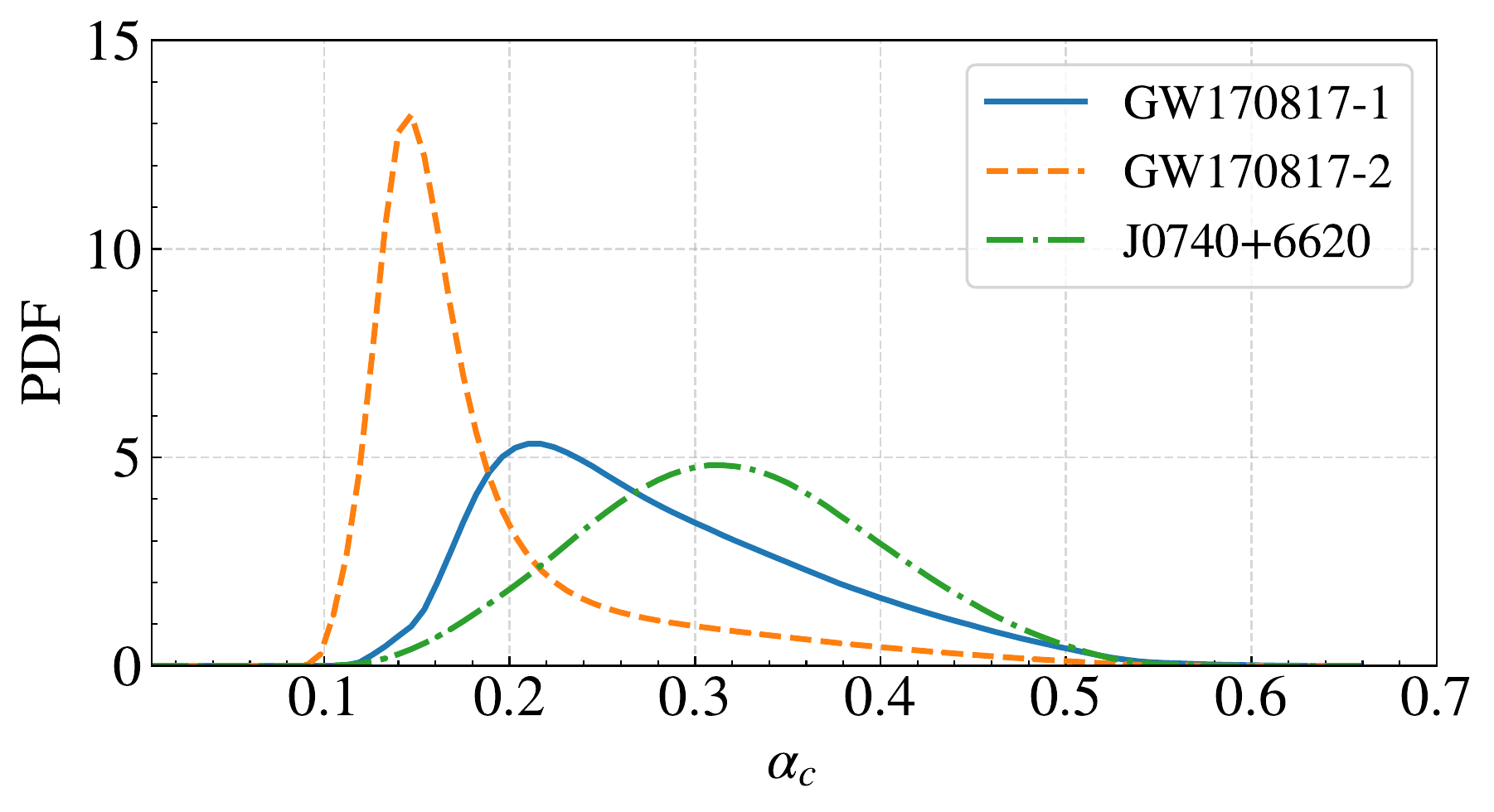}
   \caption{Posteriors for $\alpha_c$ for the primary (1) and secondary (2) components of the GW170817 event, as well as for pulsar J0740+6620.}
   \label{fig:real_data}
\end{figure}

\begin{figure}[th]
    \centering
    \includegraphics[width=0.45\textwidth]{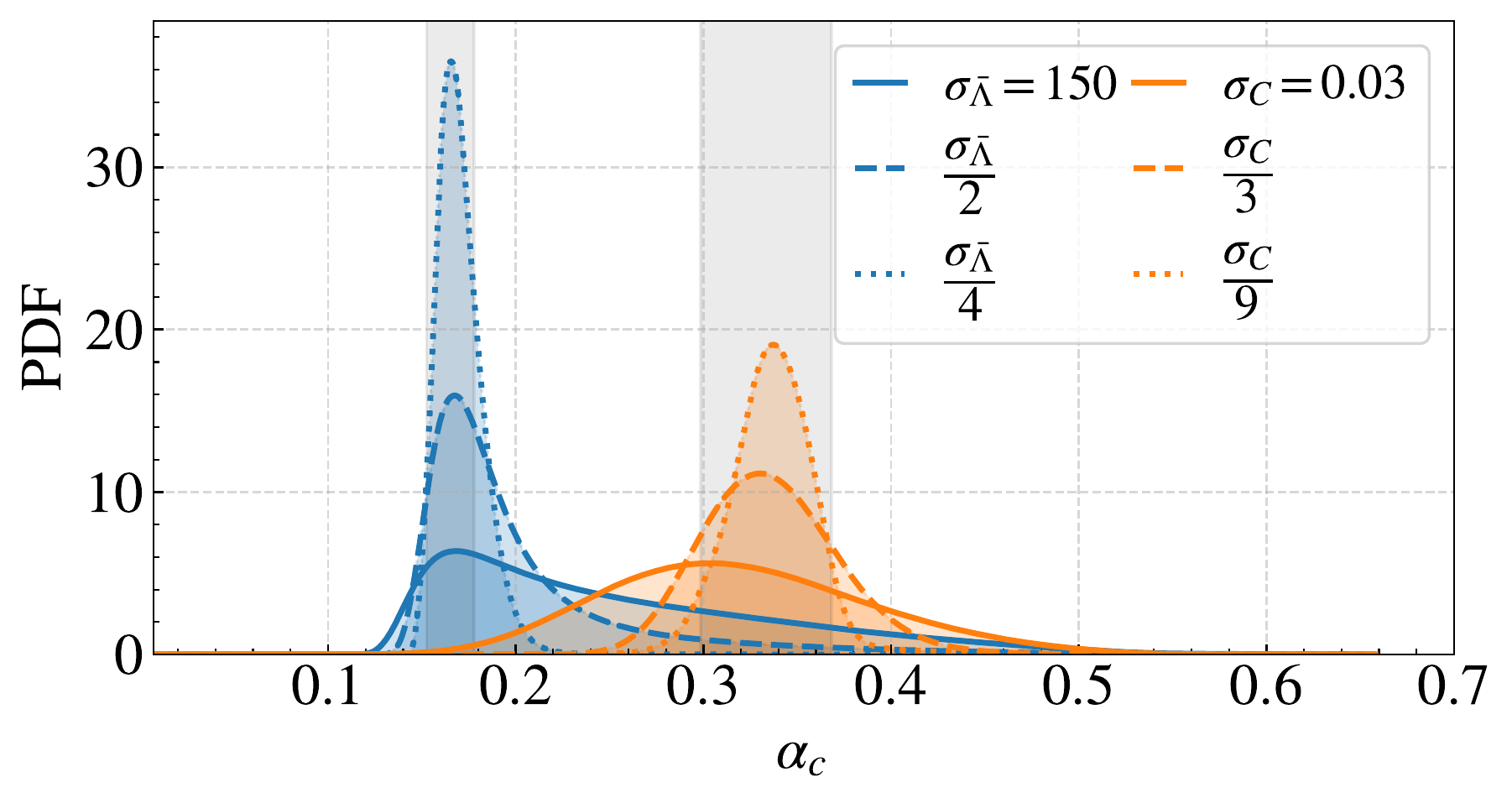}
   \caption{Posteriors for $\alpha_c$ given a hypothesized measurement of NS compactness (with mean $\mu_C = 0.26$; in blue) or dimensionless tidal deformability (with mean $\mu_{\bar{\Lambda}} = 200$; in orange). A truncated normal distribution with decreasing variance is assumed, starting with $\sigma_{C} = 0.03$ or $\sigma_{\bar{\Lambda}} = 150$. Gray vertical bands indicate 90\% credible intervals centered at the median for the underlying distribution for $\alpha_c$, inferred from Fig.~\ref{fig:universal_gpp}.}
   \label{fig:mock_data}
\end{figure}

Figure \ref{fig:real_data} shows the posterior probability density function (PDF) for $\alpha_c$ for the primary and secondary components of the binary NS system that originated GW170817, as well as for pulsar J0740+6620.
For the components of the binary system, we only use information about their tidal deformabilities ($\vec{D} = \{\bar{\Lambda}\}$) \cite{GW170817eos}, while for pulsar J0740+6620, we use information about its compactness ($\vec{D} = \{C\}$), computed from the mass and radius distributions obtained with NICER-only data by the group of Miller et al. \cite{Miller2021}. 

The spread in the PDFs in Fig.~\ref{fig:real_data} is determined by both measurement uncertainties and the residual EOS dependence of the $\alpha_c - C/\bar{\Lambda}$ relations. In order to explore the effect of increasing experimental precision (e.g.~by third-generation GW detectors \cite{Carson2019}), in Fig.~\ref{fig:mock_data} we show the posteriors for $\alpha_c$ given hypothetical observational likelihoods. 
The hypothetical measurement of observable $\mathcal{O}$ is assumed to follow a truncated ($\mathcal{O}>0$) Gaussian distribution with mean $\mu_\mathcal{O}$ and variance $\sigma_\mathcal{O}^2$. In Fig.~\ref{fig:mock_data}, $\mu_C = 0.26$ and $\mu_{\bar{\Lambda}} = 200$ and the standard deviation decreases from the initial values $\sigma_{C} = 0.03$ and $\sigma_{\bar{\Lambda}} = 150$, which are roughly compatible with GW170817 and J0740+6620 measurements. Initially, measurement uncertainties are the dominant source of error, but, as the measurement precision increases, the uncertainty in the posteriors eventually becomes dominated by the residual EOS dependence of the $\alpha_c - C/\bar{\Lambda}$ relations, which establish a lower bound for error bars (90\% credible interval) at $0.165 \pm 0.013$ (for $\bar{\Lambda} = 200$) and $0.33 \pm 0.03$ (for $C = 0.26$). This relatively small uncertainty ($\lesssim 10\%$) is a manifestation of the approximate universality of the $\alpha_c - C/\bar{\Lambda}$ relations, and contrasts with the much larger uncertainties obtained for other microscopic quantities 
--- for the same value of $C$ (resp. $\bar{\Lambda}$), the underlying distribution for $c_{s,c}^2$ has a median and 90\% credible interval of $0.66 \pm 0.19$ (resp. $0.35 \pm 0.22$). 

\section{Conclusions} \label{sec:conclusions}

In this work we have uncovered approximately universal relations between certain NS observables (such as its compactness, moment of inertia and tidal deformability) and the pressure to energy-density ratio at the center of that NS, which can be interpreted as a measure of the mean stiffness of nuclear matter inside that object (cf.~Sec.~\ref{sec:properties}). Figures \ref{fig:universal_realistic} and \ref{fig:universal_gpp} show our main results concerning the EOS-independence of the $\alpha_c - C/\bar{I}/\bar{\Lambda}$ relations, which are stronger than other known relations linking macroscopic and microscopic properties (cf.~Sec.~\ref{sec:universality_comparison}). 

It is remarkable that a microscopic quantity characterizing the core of a NS may be reasonably recovered from a single, sufficiently precise measurement of a NS property (cf.~Sec.~\ref{sec:observations}); the approximately universal $\alpha_c - C/\bar{I}/\bar{\Lambda}$ relations presented here may thus provide a window into the extremeness of nuclear matter inside NSs. 
Additionally, these relations may be useful in tests of GR. Scalar extensions of GR -- including models with cosmological applications such as the chameleon \cite{Khoury2004a} or symmetron \cite{Hinterbichler2010} -- are known to predict a different phenomenology with respect to GR for NSs with $\alpha_c > 1/3$ \cite{Mendes2016,Palenzuela2016,DeAguiar2020,Ventagli2020,Dima2021,Aguiar2021}.
Interestingly, Fig.~\ref{fig:real_data} shows that the posterior for pulsar J0740+6620 has a significant support ($\sim 43\%$ probability) in the $\alpha_c > 1/3$ region, although the precise value varies somewhat for different data sets, such as those provided by Ref.~\cite{Riley2021} or when XMM-Newton data is included in the analysis.
In any case, the strong correlation between $\alpha_c$ and $C/\bar{I}/\bar{\Lambda}$, together with more precise NS measurements, should enable the identification of systems (if these do exist in Nature) where the condition on $\alpha_c > 1/3$ is likely satisfied, opening a new window for tests of modified theories of gravity.

\acknowledgments
This work was partially supported by the National Council for Scientific and Technological Development (CNPq), and by the Carlos Chagas Filho Research Support Foundation (FAPERJ).

\bibliography{library}

\end{document}